# Operationalising Cyber Risk Management Using AI: Connecting Cyber Incidents to MITRE ATT&CK Techniques, Security Controls, and Metrics

Emad Sherif [a], Iryna Yevseyeva [a], Vitor Basto-Fernandes [b, c], Allan Cook [a]

[a] Faculty of Technology, Arts and Culture, De Montfort University, Leicester, United Kingdom
[b] Instituto Universitário De Lisboa (ISCTE-IUL), University Institute of Lisbon, ISTAR-IUL, Lisboa, Portugal
[c] Sorbonne Université, CNRS, LIP6, 75005 Paris, France

**Abstract**

The escalating frequency of cyber-attacks poses significant challenges for organisations, particularly small enterprises constrained by limited in-house expertise, insufficient knowledge, and financial resources. This research presents a novel framework that leverages Natural Language Processing to address these challenges through automated mapping of cyber incidents to adversary techniques. We introduce the 'Cyber Catalog'—a knowledge base that systematically integrates CIS Critical Security Controls, MITRE ATT&CK techniques, and SMART metrics. This integrated resource enables organisations to connect threat intelligence directly to actionable controls and measurable outcomes. To operationalise the framework, we fine-tuned all-mpnet-base-v2, a highly regarded sentence-transformers model used to convert text into numerical vectors on an augmented dataset comprising 74,986 incident-technique pairs to enhance semantic similarity between cyber incidents and MITRE ATT&CK techniques. Our fine-tuned model achieved a Spearman correlation of 0.7894 and Pearson correlation of 0.8756, demonstrating substantial improvements over top baseline models including all-mpnet-base-v2 ($\Delta\rho$ = 0.2042), all-distilroberta-v1 ($\Delta\rho$ = 0.2118), and all-MiniLM-L12-v2 ($\Delta\rho$ = 0.2309). Furthermore, our model exhibited significantly lower prediction errors (MAE = 0.135, MSE = 0.027) compared to all baseline models, confirming superior accuracy and consistency. The Cyber Catalog, training dataset, trained model, and implementation code made publicly available to facilitate further research and enable practical deployment in resource-constrained environments. This work bridges the gap between threat intelligence and operational security management, providing an actionable tool for systematic cyber incident response and evidence-based cyber risk management.

**Keywords:** Cyber Incidents, Cyber Threats, Metrics, MITRE ATT&CK, NLP, Security Controls, Sentence Transformers, SMART Criteria.

## 1. Introduction

### 1.1. Background and Motivation

The contemporary cyber threat landscape is characterised by an unprecedented escalation in both the frequency and sophistication of cyber-attacks. Organisations of all sizes face threats from advanced persistent threat (APT) groups and opportunistic attackers exploiting vulnerabilities. However, the burden of these threats falls disproportionately on small and medium-sized enterprises (SMEs), which often lack the financial resources, technical expertise, and personnel necessary to implement robust cyber security programmes.

Traditional approaches to cyber incident analysis are manual processes that demand substantial time investment from analysts. These processes involve examining unstructured incident descriptions, identifying the adversary techniques, mapping these techniques to relevant security controls, and assessing the effectiveness of implemented controls. This manual process is not only time-consuming but also prone to human error and inconsistency, particularly when dealing with the volume and variety of incidents that organisations face.

The MITRE ATT&CK framework has emerged as the de facto standard for describing adversary behaviour in the cyber security domain (MITRE, n.d.). It provides a structured taxonomy of adversary techniques based on real-world observations. Similarly, the Centre for Internet Security (CIS) Critical Security Controls offer a prioritised set of actions that collectively form a defence-in-depth security posture (Centre for Internet Security, n.d.). However, the manual process of connecting cyber incidents to ATT&CK techniques and subsequently to CIS controls remains a significant operational bottleneck.

Furthermore, whilst organisations implement various security controls, monitoring and assessing such controls often relies on subjective judgement rather than objective data. The absence of measurable indicators makes it difficult to determine whether their security measures are yielding the desired risk reduction (Savola, 2013). Research has demonstrated that effective security metrics must adhere to SMART criteria (Specific, Measurable, Achievable, Relevant, and Time-bound) to provide actionable insights for decision-makers. Without such metrics, organisations struggle to justify security expenditures, prioritise improvements, or demonstrate compliance (Hubbard & Seiersen, 2023; Sherif et al., 2024).

## 1.2. Problem Statement

Current cyber incident analysis workflows exhibit several structural limitations. Translating unstructured incident narratives into structured technique classifications (e.g., MITRE ATT&CK) consumes significant analyst time; once techniques are identified, analysts must still manually determine which security controls are pertinent, a step that presupposes fluency in both ATT&CK and control frameworks such as the CIS Critical Security Controls. Control assessment and monitoring often rely on subjective judgements rather than objective, quantifiable metrics, and manual processes fail to scale with incident volume, producing delays and potential coverage gaps. Collectively, these constraints underscore the need for automated, scalable approaches that can both match relevant adversary techniques and perform incident-to-control mapping efficiently.

Against this backdrop, the study addresses three research questions. First, it investigates the methodological challenges in constructing mappings between CIS security controls and ATT&CK techniques and asks what objective metrics can be defined to monitor those controls. Second, it evaluates the extent to which a general-purpose transformer model can accurately map cyber incidents to ATT&CK techniques and explores performance gains attainable through domain-specific fine-tuning. Third, it examines how automated incident-to-control mappings can be operationalised to support risk-based decision-making.

The work pursues three primary objectives: (1) accelerating incident triage via automated identification of relevant techniques in a globally accessible adversary knowledge base; (2) improving cyber risk management through automated linkage between incidents and security controls; and (3) enabling the assessment and ongoing monitoring of control implementation using quantifiable metrics.

The central hypothesis is that a framework which automatically maps cyber incidents to ATT&CK techniques at scale—integrated with relevant security controls assessed via objective, quantifiable metrics—will provide decision-makers with actionable intelligence for data-informed cyber risk mitigation. Such a framework should reduce risk exposure, enhance organisational cyber resilience, and strengthen the overall security posture.

### 1.3. Contribution

We make the following contributions:

1. We introduce a comprehensive knowledge base—the 'Cyber Catalog'—that systematically integrates three critical components: CIS Critical Security Controls, MITRE ATT&CK techniques, and metrics. This resource establishes mappings between security controls and adversary techniques, with each control linked to quantifiable metrics. The Cyber Catalog addresses a fundamental gap in operational cyber security by providing practitioners with a unified reference that connects security measures to threat intelligence and measurable outcomes.
2. We present a comprehensive approach for creating high-quality training data with rigorous quality assessment based on BERTScore evaluation (Zhang et al., 2020). We address the challenge of one-to-many mappings in training data through a modified loss function that prevents false negative penalties during training, combined with hard negative mining to enhance discrimination (Robinson et al., 2021).
3. We demonstrate substantial improvements achieved through fine-tuning a general-purpose sentence transformer on domain-specific cyber incident data, with improvements of approximately 0.21 in Spearman correlation over baseline models.

The remainder of the paper is organised as follows: Section 2 reviews related work. Section 3 details the methodology, including data preparation. Section 4 presents the experimental results. Section 5 discusses the implications, limitations, and potential applications of our approach. Section 6 concludes the paper.

## 2. Related Work
### 2.1. Cyber Risk Management and Threat Intelligence

Cyber risk management has evolved significantly over the years, transitioning from reactive, signature-based defence mechanisms to proactive, intelligence-driven approaches. Contemporary frameworks emphasise continuous monitoring, predictive analytics, and risk-based prioritisation. The integration of threat intelligence with security controls represents a critical capability for organisations seeking to move beyond defensive postures.

The ATT&CK framework has become the cornerstone of modern threat intelligence, providing a common language for describing adversary behaviour. Efforts have explored methods for automatically extracting techniques from unstructured text sources, including threat reports. However, most existing approaches rely on rule-based systems or traditional machine learning classifiers that struggle with the semantic nuances and contextual variations present in real-world data (Mohasseb et al., 2019; Rafiey & Namadchian, 2024).

### 2.2. Natural Language Processing

The application of NLP techniques to cyber security tasks has gained considerable attention in recent years. Early work focused on named entity recognition for extracting indicators of compromise from text (Georgescu et al., 2019). Recent research has explored document classification and semantic similarity tasks specific to the cyber security domain (Rafiey & Namadchian, 2024).

Transformer-based models, particularly BERT (Bidirectional Encoder Representations from Transformers) and its variants, have demonstrated remarkable performance across a wide range of NLP tasks (Reimers & Gurevych, 2019). Sentence-BERT and similar sentence embedding models extend these capabilities to semantic similarity tasks by producing fixed-length vector representations that capture sentence-level semantics. These models were successfully applied to tasks such as semantic search and document clustering (Cer et al., 2017; Pavlyshenko & Stasiuk, 2025).

## 2.3. Fine-Tuning

Transfer learning—the practice of adapting pre-trained models to specific domains or tasks—has become the dominant paradigm. General-purpose language models trained on large corpora can be fine-tuned on domain-specific data to achieve substantial performance improvements over baseline models (Stankevičius & Lukoševičius, 2024; Sufi, 2024). This approach is valuable in specialised domains such as cyber security, where labelled training data may be scarce or expensive to obtain.

Some research efforts have investigated fine-tuning transformer models for cyber security tasks (Bayer et al., 2023; Pavlyshenko & Stasiuk, 2025; Stankevičius & Lukoševičius, 2024). However, most existing work focused on classification rather than semantic similarity. Further, limited research has addressed the specific challenge of mapping unstructured incidents to technique taxonomies such as ATT&CK.

## 2.4. Contrastive Learning and Loss Functions

Contrastive learning frameworks, which learn representations by contrasting positive pairs against negative examples, have shown promise for semantic similarity tasks. MultipleNegativesRankingLoss (MNRL) is a widely adopted loss function that leverages in-batch negatives to improve training efficiency (Piperno et al., 2025). However, MNRL can introduce false negatives when the same target appears multiple times in a batch—a common scenario in cyber security datasets where multiple incidents may map to the same technique. Related works have explored modifications to contrastive learning frameworks to address these challenges, including hard negative mining and duplicate-aware training procedures (Cui et al., 2023; Robinson et al., 2021). These approaches are relevant for domains characterised by one-to-many relationships between input and output entities.

## 2.5. Security Control Frameworks

Various security control frameworks have been developed to provide organisations with structured guidance for implementing cyber security measures. The CIS Critical Security Controls (CIS Controls) offer a prioritised, evidence-based set of Safeguards, giving organisations a practical, essential cyber hygiene first, on-ramp before deeper measures, it comprises 18 Controls and 153 Safeguards. (Centre for Internet Security, n.d.). Other widely used frameworks provide complementary governance or sector-specific coverage: NIST CSF 2.0, NIST SP 800-53 Rev.5, ISO/IEC 27001:2022 with the restructured 27002:2022 control set, and PCI DSS 4.0. In the UK public-sector context, the NCSC Cyber Assessment Framework (CAF) sets outcome-based expectations across managing risk, protection, detection, and impact minimisation. Crucially, official CIS cross-walks align CIS Safeguards to NIST CSF, NIST 800-53, ISO/IEC 27001, and PCI DSS, allowing teams to keep a CIS-centred operational baseline while demonstrating compliance under these regimes. However, limited research has explored automated approaches for linking threat intelligence, adversary techniques, and security controls into a unified framework. Our work addresses this gap by providing an end-to-end pipeline from incident description to quantifiable metrics used to monitor control implementation.

## 3. Methodology
### 3.1. Overview

Our methodology comprises five stages: (1) creating the Cyber Catalog—a comprehensive knowledge base integrating security controls, adversary techniques, and metrics; (2) training data preparation through synthetic augmentation and quality assessment; (3) model selection and architecture configuration; (4) fine-tuning with a modified loss function to address data characteristics; and (5) comprehensive evaluation using multiple metrics and baseline comparisons. Figure 1 provides a high-level overview of our approach.

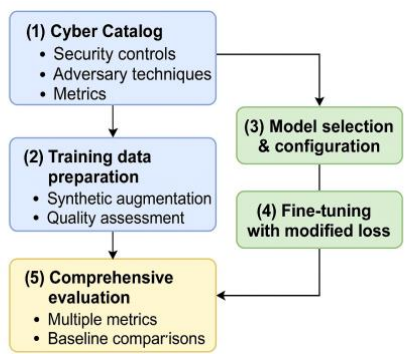

Figure 1. High-level overview of the proposed methodology.

### 3.2. Cyber Catalog Development
#### 3.2.1. Motivation and Design Principles

A fundamental challenge in operational cyber security is the disconnection between threat intelligence, defensive controls, and measurable outcomes. Security teams typically work with three distinct knowledge domains: (1) threat intelligence describing adversary behaviours (e.g., MITRE ATT&CK), (2) control frameworks prescribing defensive measures (e.g., CIS Controls), and (3) metrics for assessing control effectiveness. These domains are rarely integrated, forcing practitioners to manually bridge gaps between threat observations and defensive actions. To address this fragmentation, we created the Cyber Catalog—a unified knowledge base that systematically connects these three domains. It serves as a reference resource that enables organisations to connect threats (ATT&CK techniques) to relevant defensive controls (CIS Controls) as well as assess control effectiveness using quantifiable metrics.

#### 3.2.2. Structure and Components

The Cyber Catalog comprises three interconnected layers as outlined below.

- The foundation layer consists of the 18 CIS Critical Security Controls (v8), which represent prioritised cyber security best practices developed through community consensus. Each control addresses a specific security domain (e.g., asset management, access control, incident response) and comprises multiple safeguards that prescribe concrete implementation actions.
- The second layer incorporates MITRE ATT&CK techniques representing documented adversary behaviours. Each technique is mapped bidirectionally to relevant CIS controls, establishing explicit relationships between threats and defensive measures. These mappings enable security teams to determine which controls mitigate specific techniques and, conversely, which techniques a given control address.
- The third layer associates each security control with quantifiable metrics that enable objective assessment and

monitoring. Following established best practices in security metrics selection, all metrics adhere to SMART criteria.

Research has demonstrated that adherence to SMART criteria is essential for developing metrics that provide actionable intelligence for security decision-making. Non-SMART metrics often lead to measurement theatre—collecting data that provides little decision support value—whilst SMART metrics enable evidence-based assessment of security posture and control effectiveness.

### 3.2.3. Mapping Methodology

The mapping methodology for CIS Controls v8 to MITRE ATT&CK Techniques involves aligning each Safeguard with specific techniques and sub-techniques to illustrate how defensive actions can mitigate or detect adversarial behaviours. This mapping is done at a granular level, focusing on individual Safeguards rather than broad Controls, and uses the ATT&CK Enterprise framework. The goal is to help organizations implement threat-informed defence strategies by showing which security practices counter known attack patterns, thereby supporting prioritisation, gap analysis, and operational planning in cyber security programs. This resource was adapted from (Centre for Internet Security, 2024).

Additionally, the methodology behind the CIS Controls Assessment Specification (CAS) is designed to help organisations consistently measure the implementation of each CIS Safeguard. It distinguishes between verifying whether a Safeguard has been implemented and assessing how well it has been implemented, with the Specification focusing on the former. The approach emphasises what to measure, not how to measure, allowing flexibility across diverse technologies. Each Safeguard is broken down into structured components: assumptions, required inputs, operations to perform on those inputs, resulting measures, and metrics derived from those measures. This structure enables tool vendors and adopters to uniformly assess Safeguard implementation while accommodating technological variability (Centre for Internet Security, 2025).

The CIS Controls v8 to ATT&CK Techniques mapping is provided as a structured CSV format, facilitating automated analysis. In contrast, the CIS Controls Assessment Specification (CAS) is accessible solely via the CIS website and lacks a downloadable structured dataset. Consequently, we undertook a manual extraction and alignment process, systematically mapping each CAS component—namely Inputs, Operations, Measures, and Formula—to their corresponding Control to enable integrated assessment and analytical consistency.

### 3.2.4. Public Availability

The complete Cyber Catalog released as a structured CSV file that contains the following:

- Control descriptions (CIS v8).
- Associated ATT&CK techniques.
- Metrics, which comprises Inputs, Operations, Measures, and Formula.

By making this resource publicly available, we enable organisations of all sizes to benefit from this integration without requiring extensive expertise in multiple frameworks. The structured format facilitates integration into cyber security orchestration platforms.

### 3.3. Training Dataset Preparation
### 3.3.1. Base Dataset

The foundation of our training dataset was established using the European Repository of Cyber Incidents (EuRepoC) Global Database version 1.2, a systematically curated collection of documented cyber incidents (European Repository of Cyber Incidents, 2024). Each incident was subjected to a manual annotation process whereby cyber security domain experts mapped incident characteristics to corresponding techniques within the MITRE ATT&CK framework. This expert annotation process yielded 762 incident-technique pairs, forming the base dataset. The manual mapping performed by cyber security professionals possessing both comprehensive knowledge of the MITRE ATT&CK taxonomy and substantial expertise in analysing real-world attack patterns, thereby ensuring the reliability and validity of the annotations.

The initial dataset, whilst high-quality, was insufficient in scale for effectively fine-tuning. Deep learning models, particularly those based on transformer architectures, typically require thousands to tens of thousands of training examples to achieve optimal performance (Kaya & Bilge, 2019). This limitation necessitated a data augmentation strategy.

### 3.3.2. Synthetic Data Augmentation

According to Sufi (2024), advanced Generative Pre-Trained Transformer (GPT) architectures—such as the GPT-5 series—demonstrate superior performance in synthetic data generation tasks compared to earlier transformer models and alternative generative approaches like Generative Adversarial Networks (GANs). This improvement is attributed to enhanced model capacity, training methodologies, and contextual representation capabilities in latest transformer-based designs.

Therefore, to address the data scarcity challenge, we instructed a large language model (GPT-5) to generate synthetic incident descriptions. For each of the 762 original incidents, we instructed the model to generate 100 semantically similar but lexically distinct incident descriptions. This prompt-based augmentation approach demonstrated to be effective for creating diverse training data whilst preserving semantic content (Ding et al., 2024).

Our augmentation strategy incorporated techniques to ensure diversity and realism. For advanced paraphrasing, we employed multiple complementary techniques designed to vary linguistic expression while preserving semantic content. These techniques included synonym replacement with context-aware selection to maintain domain-specific terminology appropriateness, sentence restructuring through transformations between active and passive voice constructions, and phrase-level paraphrasing to introduce lexical variation whilst maintaining the underlying meaning of the original text. To enhance the generalisability across diverse real-world scenarios, we implemented systematic entity randomisation. This process involved varying specific entities throughout the synthetic incident descriptions, including geographic references such as countries and regions, organisation names, malware and threat actor designations, temporal references encompassing dates and time periods. This randomisation ensured that the model would not overfit to specific named entities present in the original dataset. Moreover, we enforced strict constraints to ensure semantic preservation of critical incident characteristics (Sufi, 2024). Specifically, we maintained consistency in attack vectors and techniques employed. Therefore, this systematic augmentation process generated 76,200 synthetic incident descriptions (762 original incidents times 100 augmentations per incident), yielding a larger and more diverse training corpus.

### 3.3.3. Quality Assurance and Filtering

The augmented data underwent rigorous quality assurance procedures: security experts manually reviewed the augmented dataset to verify accuracy and relevance. This review process identified and removed 1,214 duplicate or low-quality entries, resulting in 74,986 high-quality incident-technique pairs. Furthermore, we employed BERTScore, a reference-based evaluation metric that leverages contextual embeddings from BERT, to assess semantic similarity between original and augmented incidents. According to Zhang et al. (2020), BERTScore is superior to n-gram-based metrics such as ROUGE for evaluating semantic similarity because it captures meaning rather than merely surface-level lexical overlap. BERTScore computes precision, recall, and F1 scores by comparing token-level embeddings between reference and candidate sentences using cosine similarity. For our dataset, we established a threshold of 0.75 for the F1 score, below which augmented examples were excluded. The lowest scores observed in our filtered dataset are:

- Precision: 0.807 (indicating high relevance of generated content).
- Recall: 0.756 (indicating substantial capture of original content).
- F1: 0.781 (indicating strong overall semantic alignment).

These metrics demonstrate that our augmented data maintains high semantic fidelity to the original incidents whilst introducing beneficial lexical diversity.

### 3.3.4. Hard Negative Mining

A critical challenge in our training data is the one-to-many relationship between techniques and incidents: multiple incidents may map to the same ATT&CK technique. When using contrastive loss functions such as MNRL, this characteristic can introduce false negatives if multiple positive pairs sharing the same technique appear in the same batch (Cui et al., 2023). To address this limitation and improve model discrimination, we implemented a hard negative mining strategy. Hard negatives are examples that are semantically similar to the positive example but represent incorrect mappings. Including such examples during training forces the model to learn finer-grained distinctions (Robinson et al., 2021).

The hard negative mining was implemented by instructing GPT-5 to execute a systematic four-stage process. First, GPT-5 was directed to perform technique normalisation, whereby each unique ATT&CK technique description was processed through lowercasing, punctuation removal, and stop-word elimination to create standardized representations. Second, the model was tasked with computing pairwise similarity scores between all distinct techniques using a weighted combination of two complementary metrics: word-level Jaccard similarity calculated after stemming to capture lexical overlap, and character 4-gram overlap to identify morphological similarities and shared phrasal constructions. Third, GPT-5 was instructed to conduct hard negative selection by identifying, for each technique $t$, the most similar yet distinct technique $u$, where $u \neq t$, to serve as its hard negative. This selected hard negative was then systematically associated with every training instance where $t$ appeared as the positive technique, ensuring consistent pairing throughout the dataset. Finally, GPT-5 was directed to record and store the similarity score between each positive technique and its corresponding hard negative, thereby enabling potential filtering or weighting schemes to be applied during subsequent model training phases.

Therefore, this instructional approach ensures that GPT-5 generates training examples that compel the model to learn fine-grained discriminations between genuinely similar but distinct techniques, rather than relying on coarse-grained pattern recognition that might conflate related but separate attack methodologies.

### 3.3.5. Final Dataset Composition

The final training dataset contains 74,986 incident-technique pairs: 59,987 training set (80%), 7,499 validation set (10%), and 7,499 test set (10%). The test set was held out entirely during training, providing an unbiased estimate of model performance on unseen data.

## 3.4. Model Architecture
### 3.4.1. Base Model Selection

We selected all-mpnet-base-v2 as our base model. This model is a sentence transformer based on the MPNet (Masked and Permuted Pre-training for Language Understanding) architecture, which combines the advantages of BERT and XLNet whilst mitigating their respective limitations (Song et al., 2020). According to Song et al. (2020), the all-mpnet-base-v2 model generates 768-dimensional fixed-length sentence embeddings that capture semantic meaning and contextual information. Previous comparative studies have demonstrated that all-mpnet-base-v2 achieves statistically superior performance compared to other popular open-source transformers across diverse semantic similarity benchmarks.

### 3.4.2. Framework Selection

We implemented the training pipeline using PyTorch, an open-source deep learning framework renowned for flexibility, modularity, and strong support for research applications. PyTorch integrates seamlessly with the Sentence-Transformers library, which provides high-level APIs for training and evaluating sentence transformers (Hussain et al., 2025). Comparative studies have demonstrated that PyTorch achieves superior performance to TensorFlow in terms of training accuracy, precision, recall, and loss convergence for similar tasks. Furthermore, the sentence-transformers library is natively built on PyTorch, making it the natural choice for our implementation (Stanescu & Dinu, 2023).

## 3.5. Loss Function
### 3.5.1. Multiple Negatives Ranking Loss

For semantic similarity tasks, the loss function must encourage the model to produce embeddings where similar sentences are close together in the embedding space whilst dissimilar sentences are far apart. MultipleNegativesRankingLoss (MNRL) is a widely adopted loss function for this purpose (Baek et al., 2025). MNRL works with mini-batches that consist of anchor-positive pairs. For each anchor paired with its corresponding positive example. The approach considers every remaining sentence in the batch as a negative example. The function minimises the negative log-likelihood of similarity scores normalised through softmax, which effectively trains the model to assign a higher ranking to the correct positive example compared to all negative examples within the batch (Piperno et al., 2025).

### 3.5.2. Duplicate-Aware Modification

A critical limitation of MNRL in our context is the one-to-many relationship in our training data: the same technique may appear as the positive target for multiple different incidents. If two pairs sharing the same technique appear in the same mini-batch, MNRL will incorrectly treat one as a negative for the other, introducing false negative signals that can degrade model performance. To address this, we implemented a duplicate-aware modification using the NoDuplicatesDataLoader. This component ensures that no sentence string appears more than once within any given training batch. By eliminating within-batch duplicates of positive examples, we prevent the false negative problem whilst maintaining the efficiency benefits of in-batch negative sampling (Martellozzo, 2024). The duplicate-aware approach, combined with hard negative mining, provides a robust training signal that encourages fine-grained discrimination whilst avoiding penalisation of legitimate positive examples.

### 3.6. Training Configuration
#### 3.6.1. Hyperparameters

We employed standard hyperparameter values that have been validated across various sentence transformer fine-tuning tasks (Song et al., 2020), including a learning rate of $2 \times 10^{-5}$, a batch size of 16, and a total of 10 epochs. The training process incorporated a warmup period equivalent to 10% of the total training steps, with evaluation performed consistently after each epoch. This specific learning rate was selected because it provides stable optimisation without the risk of overshooting, while the batch size of 16 was chosen to represent an effective balance between ensuring gradient stability and maintaining computational efficiency within our hardware constraints.

#### 3.6.2. Computational Environment

Training was conducted on a workstation running Ubuntu 24.04 LTS, with Python version 3.12.3, Sentence-Transformers library version 5.1.2, and PyTorch version 2.9.0. The hardware used was an NVIDIA Quadro RTX 5000 GPU, which features 3,072 CUDA cores, 16GB of GDDR6 memory, and is compatible with CUDA version 12.1. The entire training process, spanning 10 epochs, took approximately 46,609 seconds to complete.

### 3.7. Evaluation Methodology
#### 3.7.1. Evaluation Metrics

We used the outlined below metrics to comprehensively assess model performance. Correlation Metrics: Semantic textual similarity tasks are typically evaluated using correlation coefficients that measure the alignment between predicted and ground-truth similarity scores (Rovetta, 2020):

Spearman Rank Correlation ($\rho$) measures monotonic relationship strength, focusing on ranking accuracy. This is the preferred metric for semantic similarity tasks as it captures the model's ability to correctly order pairs by similarity. While Pearson Correlation (r) measures linear relationship strength between predicted & true scores. According to Schober & Schwarte (2018), correlation values are typically interpreted as: very strong ($|\rho| > 0.7$), moderate ($0.5 < |\rho| \leq 0.7$), fair ($0.3 < |\rho| \leq 0.5$), or poor ($|\rho| \leq 0.3$). Further, the following metrics are widely used to assess prediction accuracy (Galli et al., 2024):

- Mean Absolute Error (MAE): The average of absolute differences between predicted and true similarity scores.
- Mean Squared Error (MSE): The average of squared differences, which penalises larger errors more heavily than MAE.

#### 3.7.2. Baseline Models

We compared our fine-tuned model, ft_mpnet_v6, against top general-purpose sentence transformer models:

- all-mpnet-base-v2: known for providing excellent embedding quality.
- all-distilroberta-v1: distilled variant of RoBERTa optimised for efficiency.
- all-MiniLM-L12-v2: compact model with 12 layers, optimised for speed.

The same held-out test set was used to evaluate our fine-tuned model against these models, using identical evaluation procedures to ensure fair comparison.

## 4. Results
### 4.1. Training Performance

Training progressed smoothly over 10 epochs, with consistent improvement in validation set performance. The duplicate-aware loss function successfully mitigated the false negative problem, as evidenced by stable convergence without the oscillations that can occur when the model receives contradictory training signals.

## 4.2. Correlation Performance

Our fine-tuned model achieved strong correlation metrics on the test set. Both metrics reveal a strong alignment between the predicted and ground-truth similarity scores. A Spearman correlation of 0.789 surpasses the 0.7 threshold for a very strong correlation, indicating that the model consistently and reliably ranks incident-technique pairs by their similarity (Schober & Schwarte, 2018). The higher Pearson correlation of 0.876 suggests that this relationship is not only monotonic but also closely approximates a linear trend.

## 4.3. Comparative Analysis: Correlation Metrics

Table 1 presents the Spearman correlation performance of our fine-tuned model compared to the top baseline models, along with the delta spearman (Δρ) representing the improvement over each model.

*Table 1. Spearman Correlation Comparison.*

| Model | Spearman (ρ) | Δρ (Improvement) |
|---|---|---|
| ft_mpnet_v6 | 0.7894 | — |
| all-mpnet-base-v2 | 0.5852 | +0.2042 |
| all-distilroberta-v1 | 0.5776 | +0.2118 |
| all-MiniLM-L12-v2 | 0.5585 | +0.2309 |

Our fine-tuned model outperforms all baseline models, with improvements ranging from 0.20 to 0.23 in Spearman correlation. These represent relative improvements of approximately 37% over the baseline models. Notably, even the base model (all-mpnet-base-v2) before fine-tuning achieves only moderate correlation (ρ = 0.585), demonstrating that general-purpose transformers, whilst capable, lack the domain-specific knowledge necessary for accurate incident-technique mapping.

## 4.4. Comparative Analysis: Error Metrics

Table 2 presents the Error metrics for our fine-tuned model and all baseline models.

*Table 2. Mean Absolute and Squared Error Comparison.*

| Model | MAE | MSE |
|---|---|---|
| ft_mpnet_v6 | 0.1352 | 0.0272 |
| all-mpnet-base-v2 | 0.5281 | 0.2859 |
| all-distilroberta-v1 | 0.4806 | 0.2417 |
| all-MiniLM-L12-v2 | 0.5663 | 0.3303 |

Our model, ft_mpnet_v6, exhibits significantly lower error rates compared to all baseline models (see Figure 2). The MAE of 0.1352 indicates on average, our model's predictions deviate from ground truth by only 0.1352 units on the similarity scale. This represents approximately 67% reduction in MAE compared to the baseline models. The MSE results tell a similar story: our model achieves an MSE of 0.027, compared to 0.2859–0.3303 for baseline models. The lower MSE indicates that our model not only has lower average error but also produces fewer large errors—an important characteristic for practical deployment where outlier predictions can have significant consequences.

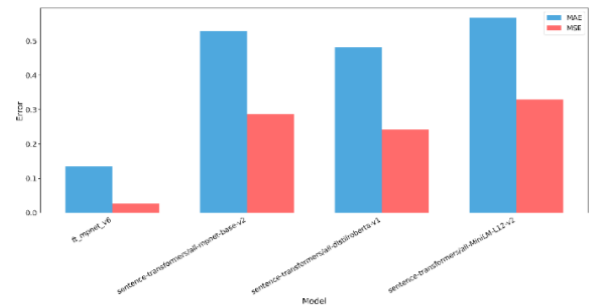

*Figure 2. MAE and MSE Comparison.*

## 4.5. Error Distribution Analysis

To gain deeper insight into model behaviour, we examined the distribution of absolute errors across the test set. Figure 3

illustrates the error distributions for all models.

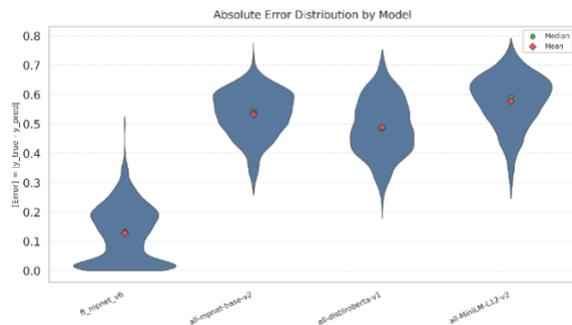

*Figure 3. Violin plot of absolute error distribution.*

Several key observations emerge from the analysis of model performance distributions. Our model, ft_mpnet_v6, exhibits the lowest median error, as indicated by the thick horizontal bar within the violin plot, thereby confirming superior average-case performance relative to baseline approaches. Moreover, the distribution for our model is notably narrow and tightly concentrated near zero, indicating consistent accuracy across the entire test set rather than performance that varies between different instances. Importantly, our model exhibits minimal probability density in high-error regions, whereas baseline models demonstrate non-negligible probability mass at large error values, suggesting that these baselines are susceptible to occasional severe mispredictions. Collectively, this narrow error distribution demonstrates that our model is not only more accurate on average but also exhibits greater stability and predictability in its behaviour, a critical characteristic for deployment in operational cyber security contexts where reliability is paramount.

# 5. Discussion
## 5.1. Interpretation of Results

The experimental results demonstrate that domain-specific fine-tuning yields substantial improvements over general-purpose sentence transformers for the task of mapping cyber incidents to MITRE ATT&CK techniques. The Spearman correlation improvement of approximately 0.2 over baseline models represents a meaningful practical enhancement: the difference between a model that provides useful similarity rankings ($\rho \approx 0.59$) and one that provides reliable, actionable rankings ($\rho \approx 0.79$).

Furthermore, the error analysis provides additional insight into model behaviour. The three-fold reduction in MAE compared to top baseline models indicates that our fine-tuned model's predictions are closer to ground truth. More importantly, the narrow error distribution demonstrates consistency—a critical property for building trust in automated systems that support decision-making.

## 5.2. Impact of Training Data Quality

The performance gains are attributed to several methodological choices: The use of LLM-based data augmentation allowed to expand the dataset (762 pairs) into a large training corpus (74,986 pairs) whilst maintaining semantic quality. BERTScore ensured that only high-quality examples are retained, preventing the model from learning spurious patterns from poorly generated data. Moreover, the inclusion of hard negatives—semantically similar but incorrect techniques—forced the model to learn fine-grained distinctions rather than merely broad categorical differences. This is used to address false negatives arising from one-to-many mappings prevented the model from receiving contradictory training signals. This contributed to more stable and effective learning.

## 5.3. Practical Applications

The developed model, ft_mpnet_v6, combined with the Cyber Catalog, enables immediate practical applications as outlined below.

Security analysts can use the integrated framework to automatically map incidents to relevant ATT&CK techniques and subsequently to applicable CIS controls. This end-to-end automation dramatically reduces the time required for incident triage and response planning. The fine-tuned model maps relevant ATT&CK techniques, enabling rapid classification even during high-volume incident scenarios. This acceleration in speed is particularly advantageous in organisations facing resource limitations and having finite analyst capacity. By mapping incidents to techniques and subsequently to CIS controls via the Cyber Catalog, the framework enables organisations to identify gaps in their security posture. If incidents consistently map to techniques for which relevant controls are not implemented, this signals a prioritisation opportunity for investments.

Moreover, the Cyber Catalog's integration of metrics enables objective, quantifiable assessment of control effectiveness. Rather than relying on subjective judgement or compliance checklists, organisations can track metrics over time to determine whether implemented controls are reducing relevant incidents and achieving desired security outcomes. This evidence-based approach supports resource allocation decisions and demonstrates security programme value to stakeholders. The automated mapping enables data-driven risk analysis. By aggregating incident patterns over time and correlating them with control implementation status and metrics, organisations can identify which techniques pose the highest threat to their environment, which controls provide the highest risk reduction, and where to allocate resources for maximum impact. The structured mappings in the Cyber Catalog facilitate compliance reporting by connecting controls to potential threats and measurable outcomes. Furthermore, the Cyber Catalog serves as an educational resource for less experienced security practitioners, helping them understand relationships between threats, controls, and metrics. This supports organisations with limited expertise.

## 5.4. Addressing the Research Questions

We have demonstrated that sentence transformer models, when fine-tuned on domain-specific data, can facilitate accurate automated correspondence between cyber incidents and security controls via MITRE ATT&CK techniques.

RQ1: Key challenges include: (1) obtaining sufficient high-quality training data, which we addressed through synthetic data augmentation; (2) handling one-to-many mappings, which we addressed through duplicate-aware training; and (3) achieving fine-grained discrimination between similar techniques, which we addressed through hard negative mining. Objective metrics based on model prediction accuracy (MAE & MSE) and rank correlation (Spearman $\rho$) provide quantifiable assessments of model performance.

RQ2: General-purpose transformer models achieve moderate performance ($\rho \approx 0.59$) on this task, indicating that they capture semantic patterns relevant to incident-technique mapping. However, domain-specific fine-tuning (our fine-tuned model) yields substantial improvements ($\Delta\rho \approx 0.21$), elevating performance to the very strong correlation regime.

RQ3: The automated mappings enable decision-support capabilities, such as prioritisation of control implementation based on observed incident patterns, objective assessment of control effectiveness through metrics, and identification of security posture gaps.

## 5.5. Limitations

The cyber security landscape evolves rapidly, with new attack techniques emerging regularly. Our model trained on historical incidents and may require periodic retraining to maintain accuracy as the threat landscape changes. The model was trained on English incident descriptions. We did not consider incidents described in other languages; this would require multilingual models or translation pipelines. Moreover, the successful integration of these models into operational workflows necessitates the establishment of robust governance protocols. Specifically, organisations should implement confidence thresholds below which automated mappings are automatically flagged for manual adjudication.

# 6. Conclusion

This research has presented a comprehensive framework for automating the mapping of cyber incidents to MITRE ATT&CK techniques and subsequently to actionable security controls using a fine-tuned sentence transformer model, ft_mpnet_v6. A central contribution of this work is the Cyber Catalog—a novel knowledge base that systematically integrates CIS Critical Security Controls, MITRE ATT&CK techniques, and quantifiable metrics. This integrated resource addresses a fundamental gap in operational cyber security by enabling organisations to translate threat intelligence into defensive actions and measurable outcomes. Through rigorous methodology encompassing data augmentation with quality controls, duplicate-aware contrastive learning, and hard negative mining, we achieved substantial performance improvements over top baseline models. Our fine-tuned model attained a Spearman correlation of 0.789 and Pearson correlation of 0.876, representing improvements of approximately 0.21 over baseline models.

Furthermore, our model, ft_mpnet_v6, exhibited significantly lower prediction errors (MAE: 0.135 & MSE: 0.027)—these metrics demonstrated three-fold reductions in MAE and MSE compared to the top baseline models, with a narrow error distribution confirming consistent, reliable predictions across diverse incident types. The framework addresses critical operational challenges in cyber risk management: accelerating incident triage, enabling security control gap identification, facilitating objective control effectiveness assessment through quantifiable metrics, and supporting data-driven risk management. By bridging the gap between unstructured threat intelligence and structured security controls with objective metrics, this work provides practical value to organisations—particularly resource-constrained SMEs—seeking to strengthen their cyber security posture through evidence-based decision-making.

In prospective research endeavours, the objective is to synthesize the fine-tuned model within an open-source host intrusion detection system. This will be achieved through the engineering of a specialised plugin, designed to leverage the capabilities of the Cyber Catalog for real-time analysis and response, operationalising dynamic, threat-informed defense mechanisms. Beyond this targeted integration, a future direction involves the strategic expansion of the Cyber Catalog's scope. This expansion will specifically incorporate comprehensive mappings to the security controls articulated within NIST Special Publication 800-53, thereby enhancing the framework's versatility and utility. This will significantly broaden the framework's applicability, providing robust support for organisations mandated to align their cyber security posture with diverse and stringent regulatory and compliance standards.